\documentclass[aps,english,prl,amsmath,amsfonts,amssymb,superscriptaddress,twocolumn,showpacs,citeautoscript]{revtex4-1}
\usepackage[T1]{fontenc} \usepackage[latin9]{inputenc}
\usepackage{amsmath} 
\usepackage{amssymb}
\usepackage{wasysym} 
\usepackage{esint} 
\usepackage{graphicx}
\usepackage{times} 
\usepackage{nicefrac} 
\usepackage{appendix}
\usepackage{textcase} 
\usepackage{subfigure} 
\usepackage{empheq}
\usepackage{color,soul} 
\usepackage{leftidx} 
\usepackage{makecell}
\usepackage{stackrel} 
\makeatletter

\renewcommand{\eqref}[1]{(\ref{eq:#1})}

\begin{document}
\title{Inverse Design of Compact Multimode Cavity Couplers}

\author{Weiliang Jin}
\affiliation{Department of Electrical Engineering, Princeton University, Princeton, NJ, 08544}

\author{Sean Molesky}
\affiliation{Department of Electrical Engineering, Princeton University, Princeton, NJ, 08544}

\author{Zin Lin}
\affiliation{John A. Paulson School of Engineering and Applied Sciences Harvard University, Cambridge, MA, 02138}

\author{Kai-Mei C. Fu}

\affiliation{Department of Physics University of Washington, Seattle, WA 98195}

\author{Alejandro W. Rodriguez}
\affiliation{Department of Electrical Engineering, Princeton University, Princeton, NJ, 08544}
\email{arod@princeton.edu}

\begin{abstract}
  Efficient coupling between on-chip sources and cavities plays a key
  role in silicon photonics. However, despite the importance of this
  basic functionality, there are few systematic design tools to
  simultaneously control coupling between multiple modes in a compact
  resonator and a single waveguide. Here, we propose a large-scale
  adjoint optimization approach to produce wavelength-scale
  waveguide--cavity couplers operating over tunable and broad
  frequency bands. We numerically demonstrate couplers discovered by
  this method that can achieve critical, or nearly critical, coupling
  between multi-ring cavities and a single waveguide at up to six
  widely separated wavelengths spanning the $560$--$1500$~nm range
  of interest for on-chip nonlinear optical devices.
\end{abstract}
\maketitle

Practical limitations of nanophotonics for broadband applications are
seldom known and highly context
specific~\cite{Yu10,miller2015shape,arabi17}.  In single frequency
problems, traditional design principles based on index-guiding, Bragg
scattering, and material resonances offer clear trade offs, e.g.,
spatial confinement for radiative losses, or peak performance for
bandwidth. But, in many multi-frequency problems, including nonlinear
optics~\cite{sitawarin2017inverse,li2016efficient},
imaging~\cite{sell2017large,callewaert2018inverse}, radiative heat
transfer~\cite{jin2017overcoming,jin2018frequency}, and optical
networks~\cite{miller2017silicon}, the number of
parameters that influence performance is simply too large to treat
completely using analytic methods or hand-designed geometries, and it
is often unclear what level of performance can be attained.  Over the
last two decades, this challenge has spurred the development of
large-scale optimization (inverse) techniques to assist in the design
process with promising early
returns~\cite{jensen2011topology,jlu_oe2013,lalau-keraly_oe2013}. Yet,
in many technically important areas only preliminary investigations
have been made~\cite{molesky2018outlook}. Specifically, the power required to attain efficient nonlinear frequency conversion processes is known to decrease with
increasing spatial confinement, so long as the overlap of the
participating modes can be controlled~\cite{rodriguez2007chi}. To take
full advantage of this effect, we have recently proposed several
optimized resonators to enhance nonlinear second harmonic and
difference frequency generation in wavelength-scale
volumes~\cite{lin2016cavity,lin2017topology}. To operate on-chip, each
mode in a device making use of these cavities must be coupled to a
source or detector in a controlled way; and until presently, we have
not addressed how this can be done. Using a typical evanescent scheme,
tuning the gap separation to control evanescent overlap between the
waveguide and cavity~\cite{chandrahalim2017evanescent}, realization of
high efficiency devices using these cavities, and similar future
designs, may be difficult. Beyond the issues of layout intricacy,
bending loss~\cite{vlasov2004losses,fujisawa2017low}, and waveguide
crosstalk~\cite{donzella2013study,jahani2015photonic} that would be
introduced by requiring multiple waveguides to intersect in a
wavelength-scale area, modes in the best performing cavities designs
may be tightly confined to the core~\cite{lin2017topology}, precluding
the possibility of achieving critical or over coupling by simply
decreasing the separation.

This problem of efficiently coupling light between sources and
predefined volumes appears in many branches of nanophotonics. For
instance, it is the defining goal of wide-area absorbers---surfaces
that can perfectly absorb a wide range of incident propagating waves.
Broadly, the main approach in this setting is to create structures
supporting many resonances in order to tune the radiative and
absorptive decay rates in each scattering
channel\cite{ghebrebrhan2011tailoring}.  This behaviour can be
introduced in a wide variety of ways, including adiabatic
tapers~\cite{wu2016broadband,lin2017tungsten},
metasurfaces~\cite{argyropoulos2013broadband,liu2017experimental},
epsilon-near-zero thin
films~\cite{molesky2013high,rensberg2017epsilon}, chirped
gratings~\cite{song2013near,ji2014broadband}, multi-resonant photonic
crystals~\cite{rinnerbauer2014superlattice}, and more recently,
unintuitive structures obtained via inverse
design~\cite{ganapati2014light,sui2015topology,fu2016broadband}. A
similar objective also appears in the context of free-space to on-chip
couplers, with the primary aim being to reduce losses, i.e.
reflections, of light incident on a on-chip device from either a fiber
or free space. Rate matching is more difficult to implement in these
situations, as any signal decay (e.g. material absorption in the
coupling region) reduces performance; and common approaches based on
adiabatic tapers lead to couplers that are several wavelengths long
and are only typically designed to operate over narrow, selective
bands~\cite{almeida2003nanotaper,carmon2007wavelength,fu2014efficient,tiecke2015efficient}.
Based on motivations similar to those of this present study, there is
a current push to exploit inverse
design~\cite{niederberger2014sensitivity,mansoor2017optimization},
metasurface concepts~\cite{zhu2017very,li2017controlling,li20182}, and
chaotic deformations~\cite{jiang2017chaos} in this area. Likewise, a
need to control coupling arises between on-chip devices, including
filters, rectifiers, multiplexers, and frequency converters.  In these
situations, the usual goal is to efficiently couple two or more separately
designed devices in the smallest possible footprint.  Again, much in
the spirit of the results presented here, within the past few years
inverse design approaches have started to be applied in this setting,
leading to experimental demonstrations of compact wavelength-division
mulitplexers operating over several far-apart
wavelengths~\cite{piggott2015inverse,frellsen2016topology}.

In this paper, we present a large-scale optimization algorithm for
designing compact on-chip devices that efficiently couple light
consisting of multiple, widely separated wavelengths from a single
waveguide into a wavelength-scale multi-resonant cavity. Motivated by
practical problems in nonlinear optics, we pursue three illustrative
examples: compact multi-resonant cavities with resonant features
mimicking those used for second-harmonic, sum-frequency, and
frequency comb generation.  In each situations, we
demonstrate either total or near total critical coupling.

\begin{figure}[t!]
  \centering \includegraphics[width=0.6\columnwidth]{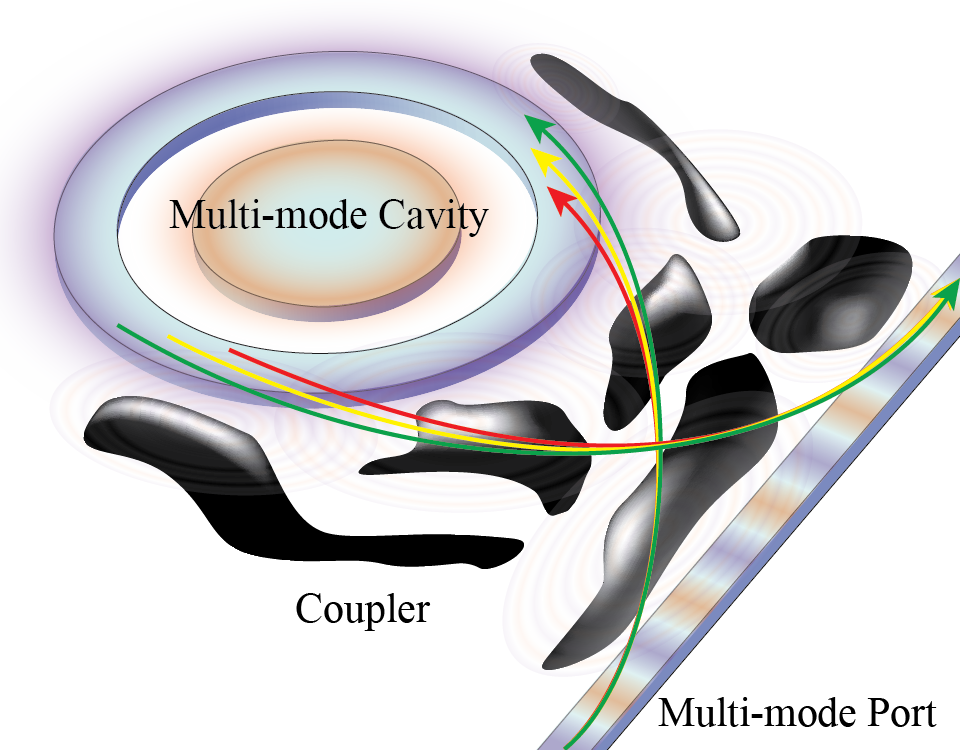} 
  \caption{\textbf{Schematic of a general cavity coupler}: A compact
    scatter (black region) acts as a coupler between a
    wavelength-scale, multimode cavity and a multimode port
    (waveguide). The design freedom of the scatter enables
    controllable coupling between the two devices at several
    wavelengths (red, yellow, and green arrows).}
  \vspace{-10 pt}
\end{figure}

\emph{Formulation.---} Our conception of the coupling problem is
depicted in Fig.~1. Starting from an isolated cavity supporting $N$
resonances with frequencies $\omega_i$ and radiative lifetimes
$Q^0_{i,r}$, $i = \{1,2,\cdots, N\}$, we aim to design a
wavelength-scale device that tunes the external coupling rate of each
mode to a single nearby waveguide to any desired value. That is, we
seek to independently control the dimensionless coupling quality
factor $Q_{i,c}$ of every individual mode of a given set. Generically,
the presence of a coupler or a nearby waveguide can significantly
alter the radiative decay of an isolated cavity modes, either
enhancing or degrading temporal confinement. To ensure that the
resonant features of the cavity are not destroyed by the coupler, we
simultaneously constrain $Q_{i,r}^0\leq Q_{i,r}\alpha_{i}$, with
$Q_{i,r}$ denoting the radiative quality factor of the cavity in the
presence of the waveguide and coupler, and $\alpha_{i}$ an adjustable
scale factor.  Based on this description, a structure for any desired
collection of coupling characteristics is discoverable using a
gradient adjoint-variable topology optimization
approach~\cite{liang2013formulation} that seeks to solve the minimax
problem,
\begin{align}
  &\min_{\{\bar{\varepsilon}\}} \mathcal{F}(\mathbf{E},\bar{\varepsilon})\\
  \text{s.t.}\;\;&\{
  \mathcal{G}_i(\mathbf{E},\bar{\varepsilon})\leq 0\},\;
  i=1,2,\cdots, N\nonumber\\
  & \varepsilon_{\text{sub}}\leq \bar{\varepsilon}\leq
  \varepsilon_{st},\nonumber
\end{align}
where
\begin{align}
  &\mathcal{F}(\mathbf{E},\bar{\varepsilon})= \max_{i=1}^N\left[ Q_{i,c}(\mathbf{E},\bar{\varepsilon})-\xi_iQ_{i,r}(\mathbf{E},\bar{\varepsilon}) \right]^2
  \label{eq:constraints}\\
  &\mathcal{G}_i(\mathbf{E},\bar{\varepsilon})=Q_{i,r}^0-\alpha_iQ_{i,r}(\mathbf{E},\bar{\varepsilon}),\nonumber
\end{align}
with $\xi_i$ denoting the target ratio of $Q_{i,c}/Q_{i,r}$. In this
method, the dielectric permittivity at every spatial point inside the
coupling region, $\{\bar{\varepsilon}\}$, is as a continuous degree of
freedom, bounded by the substrate $\varepsilon_{\text{sub}}$ and
structure $\varepsilon_{\text{st}}$ materials. (To produce binary,
smooth, fabricable systems additional regularization and filter
projection steps are applied in conjunction with this base
algorithm~\cite{wang2011projection}.) In order to circumvent numerical
issues associated with optimizations of electromagnetic
eigenvalues~\cite{liang2013formulation}, each $Q$ is computed by
solving a set of scattering problems. This makes both the objective
$\mathcal{F}$ and constraints $\mathcal{G}_{i}$ explicit functions of
the electric field $\mathbf{E}$, computed as the solution of the
steady-state equation
$\left[ \nabla\times
  \frac{1}{\mu}\nabla\times-\omega_i^2\varepsilon(\omega_i,\mathbf{r})\right]\mathbf{E}(\omega_i,\mathbf{r})=i\omega
\mathbf{J}(\omega_i,\mathbf{r})$.  To setup this problem, electric
current sources, the duals
$\mathbf{J}\left(\omega_{i}\right)\propto\text{Re}\left[\epsilon\left(\omega_{i}\right)\right]
\mathbf{E}^{*}\left(\omega_{i}\right)$ of the modes in the energy
norm~\cite{chembo2010modal}, are first calculated (without the
waveguide and coupler) at each individual frequency. The waveguide and
coupler are then added, and the field quantities of interest
determined: the electromagnetic energy density inside the cavity
volume,
$U_i=\frac{1}{2}\int_{V}\text{d}V\varepsilon(\omega_i,\mathbf{r})|\mathbf{E}(\omega_i,\mathbf{r})|^2$,
and the Poynting flux into the waveguide and radiated into vacuum,
$P_{i}=\frac{1}{2}\int_{\Sigma}\text{d}\mathbf{s}\cdot\mathrm{Re}[\mathbf{E}(\omega_i,\mathbf{r})^{*}\times
\mathbf{H}(\omega_i,\mathbf{r})]$, with $\Sigma$ denoting the
corresponding flux surfaces. The radiative and coupling lifetimes
$Q_{i,c(r)}=\omega_iU_i/P_{i,c(r)}$ are then used to evaluate
$\mathcal{F}$ and $\mathcal{G}_{i}$.

In many applications of interest, one of two coupling characteristics
are often desired: over
coupling~\cite{guo2016chip,pfeiffer2017coupling}, minimizing unwanted
losses and increasing energy efficiency, or critical coupling,
maximizing field amplitudes in the cavity~\cite{lin2016cavity}. For
cavities designed to enhance nonlinear frequency conversion, such as
the illustrative examples considered below, maximum power conversion
occurs under critical coupling, $\xi_i=1$, at each frequency. In such
cases, the general scheme presented above can be simplified. When the
cavity is pumped from a single channel with power $P^{\text{in}}$, the
energy in the cavity is related to the quality factors
by~\cite{suh2004temporal}
\begin{equation}
\frac{\omega_i\;U_i}{P^{\text{in}}_i}=\frac{4\;Q_{i,r}}{
  2+Q_{i,r}/Q_{i,c}+Q_{i,c}/Q_{i,r}},
\label{eq:U}
\end{equation}
reaching a relative maximum of $Q_{i,r}$ as the system moves toward
critical coupling ($\xi_i\rightarrow 1$). Technically, \eqref{U} is
only applicable to unidirectional couplers, i.e. when each cavity mode
couples only to one port (direction) of the waveguide. (A simple
example of a unidirectional coupler is the usual waveguide--ring
resonator system, where the direction of coupling is constrained by
momentum conservation~\cite{gorodetsky1999optical}.) However, as the
introduction of any additional coupling channel always reduces the
energy stored in the cavity~\cite{suh2004temporal}, in practice, there
is no loss of generality in considering this expression. By maximizing
the energy in the cavity, the algorithm naturally proceeds towards
unidirectional couplers, which in turn makes \eqref{U} an increasingly
good approximation. Since the behavior of \eqref{U} is then ultimately
equivalent to the more complicated \eqref{constraints}, we are able to
consider the simpler optimization problem,
\begin{equation}
\mathcal{F}'(\bar{\varepsilon})=\max_{\bar{\varepsilon}}\left\{\min_{i=1}^N \left[\frac{U_i(\mathbf{E},\bar{\varepsilon})}{U_i^0} \right]\right\}.
\label{eq:Uopt}
\end{equation}
Where the $U_i^{0} = Q_{i,r}^0 P_i^{\text{in}}/\omega_i$ is an energy normalization factor given by the bare radiative lifetime.
\begin{figure*}[t!]
  \centering
  \includegraphics[width=1.8\columnwidth]{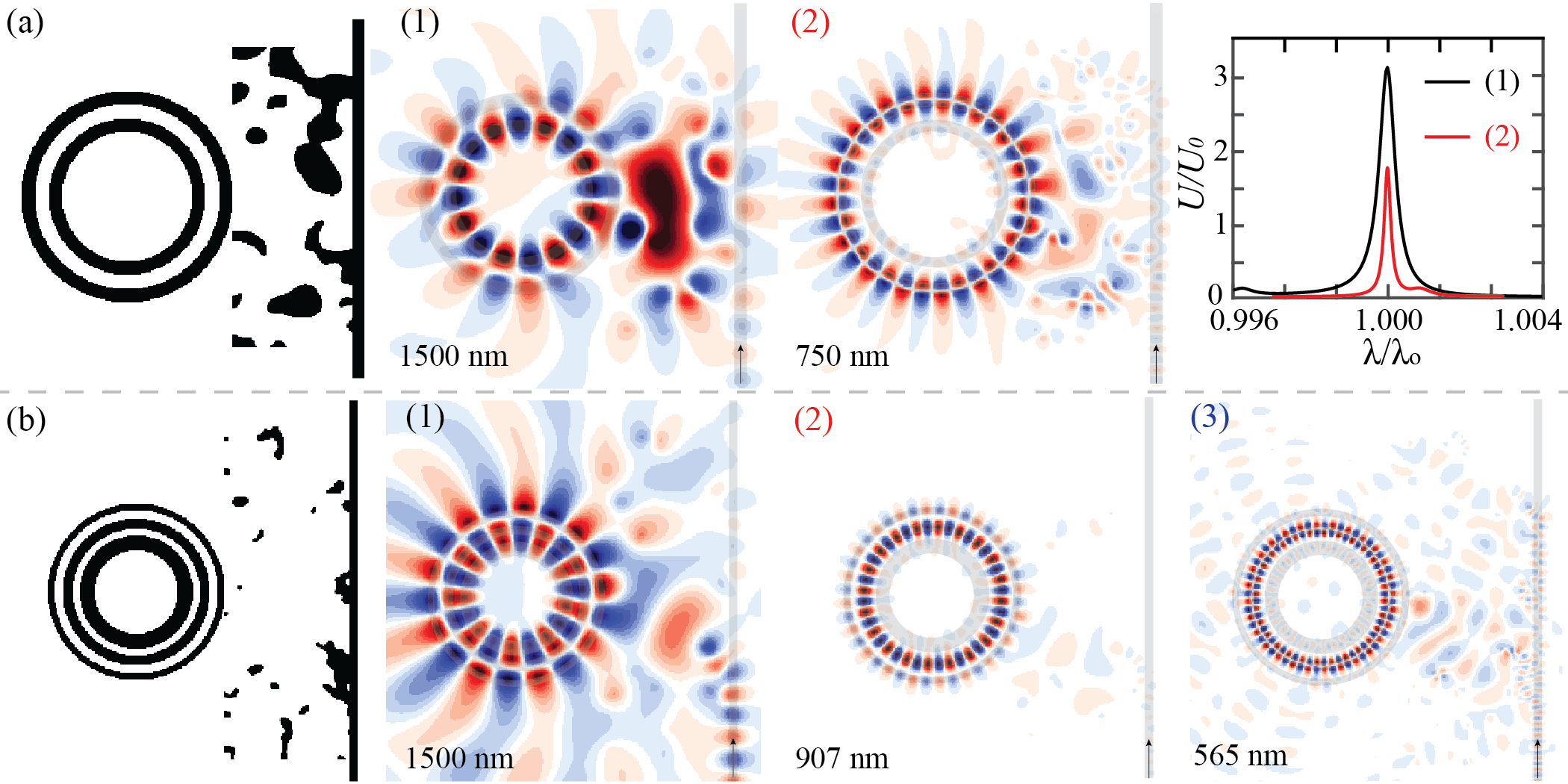}
  \vspace{0 pt}
  \caption{\textbf{Nonlinear frequency conversion}: Optimized couplers
    for SHG (a) and SFG (b) showing critical coupling between
    multimode ring resonators and waveguides. All structures (black)
    are made of GaP, while the substrate (white) is assumed to be
    vacuum. For the SHG design, the width of the waveguide is
    $150$~nm, the diameter of the outer ring 2.6 $\mu$m, and the area
    of the designed coupling region 3.75 $\mu$m $\times$ 1.5
    $\mu$m. The plot to the far right shows the energy spectrum inside
    the resonator near the fundamental and second-harmonic wavelengths
    $\lambda_{\left\{1,s\right\}}=\left\{1500,\;750\right\}$~nm, with
    matched azimuthal wavenumbers $m_1=8$ and $m_s=2m_1$, (black, red)
    normalized by $U^0$.  The middle
    figures show the TM-polarized electric fields at the respective
    wavelengths. The complete suppression of outgoing/transmitted
    power through the waveguide provides a visual confirmation of
    critical coupling. Similar results are seen for the SFG design, (b),
    with three modes $\lambda_{\{1,2,s\}}=\{1500,\;907,\;565\}$~nm,
    $m_{\{1,2,s\}}=\{9,\;20,\;28\}$, critically coupled between the
    cavity and waveguide. In this case, the width of the waveguide is
    $134$~nm, the diameter of the outer ring 2.8 $\mu$m, and the area
    of the coupling region is 5.4 $\mu$m $\times$ 2 $\mu$m. In both
    designs, the discovered structures are binary.}  \vspace{0 pt}
\end{figure*}
As a proof of concept, we consider two illustrative cavities designed
to enhance two $\chi^{(2)}$ nonlinear processes: up-conversion of
$\omega_1$ and $\omega_2$ to the summed frequency
$\omega_s=\omega_1+\omega_2$ (SFG), and second-harmonic generation
(SHG) corresponding to degenerate SFG with $\omega_1=\omega_2$. For
these processes, the relative coupling rates largely dictate the
achievable intensities in the cavity, and hence power requirements (in
the undepleted regime~\cite{lin2016cavity}). Mathematically, this is
captured by the figure of merit
\begin{equation}
  \text{FOM} = |\beta|^2\prod_{i=1,2,s}\frac{Q_{i,r}}{2+ Q_{i,r}/Q_{i,c}+Q_{i,c}/Q_{i,r}},
  \label{eq:FOM}
\end{equation}
with $\beta$ denoting the overlap coefficient of the cavity fields,
which to first order is not affected by the external waveguide or
coupler. Like \eqref{U}, \eqref{FOM} is maximized when all three modes achieve
critical coupling, giving
$\text{FOM}_\text{max} = |\beta|^2Q_{1,r}Q_{2,r}Q_{s,r}/64$.
\begin{figure*}[t!]
  \centering
  \includegraphics[width=1.8\columnwidth]{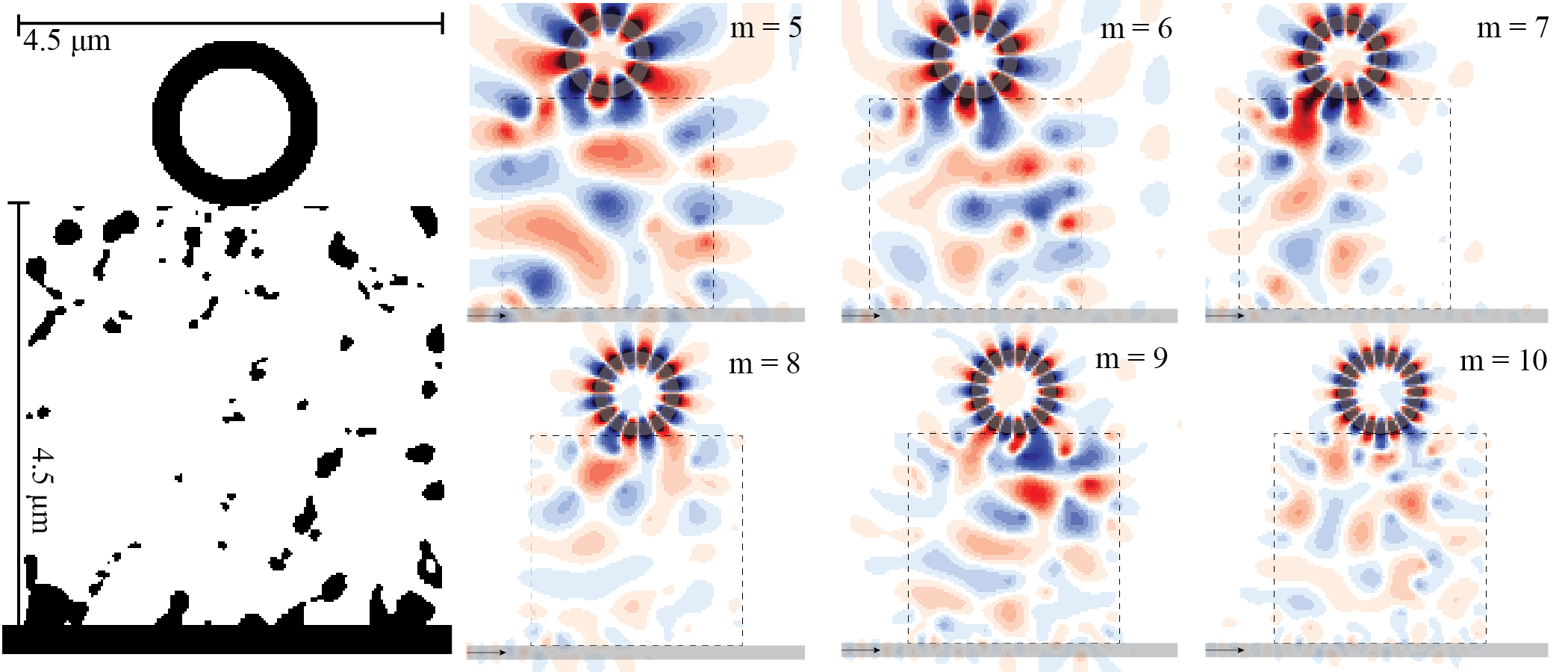}
  \vspace{0 pt}
  \caption{\textbf{Frequency comb generation}: Optimized coupler for
    comb generation showing critical coupling over 6 frequencies. The
    width of the waveguide is $300$~nm, the diameter of the ring
    $1.8\;\mu$m, and the area of the desgined coupling region $4.5$
    $\mu$m $\times$ $4.5$ $\mu$m. The figures show the TM-polarized
    electric field profiles at the respective azimuthal number
    $m=\{5\rightarrow 10\}$, corresponding to frequencies
    $f=\{0.667\rightarrow 1.157\}$ c/$1.5\;\mu$m, with equal spacing
    $\Delta f=0.098$ c/$1.5\;\mu$m.}
  \vspace{0 pt}
\end{figure*}

\emph{Results.---} As a platform for testing our algorithm, similar to
proposed wavelength-scale cavities for implementing nonlinear
phenomena\cite{lin2017topology}, we consider a two-dimensional system
consisting of hand-designed multi-track ring-resonators supporting
TM-polarized resonances of moderate radiative lifetimes
$Q_{i,r}^0\lesssim 10^5$, and a rectangular admissible coupler region
covering the separation between the cavity and the waveguide.  The
size of this design region is determined on a case-by-case basis as a
compromise between compactness and functionality.  Starting from a
base of 3.75 $\mu$m $\times$ 1.5 $\mu$m the size of the coupling
region is increased whenever the algorithm is unable to find suitable
coupling structures. To guide the algorithm towards more easily
fabricable structures, the coupler is always seeded with connected,
smooth dielectric profiles, i.e. a random ribbon. (For numerical
accuracy, the grid resolution is chosen to be smaller than
$\lambda / 45$ for the smallest wavelength considered, $\approx 13$
pixels per wavelength inside the highest index media.)

Our findings begin with the SHG and SFG systems depicted Fig.~2. For
practical considerations, these simulations suppose a gallium
phosphide material (including material dispersion) for all dielectric
regions: the cavity, waveguide, and coupler.
The initial SHG system is a two-track multi-ring
supporting TM-polarized resonances at
$\lambda_{\left\{1,s\right\}}=\left\{1500,\;750\right\}$~nm, quality
factors of $Q^0_{\{1,s\},r} = \{1.4,4.6\}\times 10^3$, with power
coupled into the device through a narrow waveguide at a gap separation
of 1.5$\lambda_1$. (The azimuthal numbers of these modes satisfy the
phase-matching condition $m_s=2m_1=16$ for the (111) plane of a GaP
crystal. Given a different nonlinear tensor and requisite
polarizations, the phase-matching condition for $m$ can be slightly
different~\cite{bi2012high}.)  As indicated in the field profiles of
Fig. 2 (a), in the presence of the coupler each mode shows vanishing
transmission and reflection ($\lesssim 2\%$), and large field
amplitude inside the cavity.  Quantitatively, Fig. 2 (a) (rightmost)
examines the energy spectrum inside the resonator channeled from the
waveguide around $\lambda_{1,2}$, normalized by $U_i^0$. After
optimization, the cavity mode lifetime is more than doubled, with
$Q_{\left\{1,s\right\},r}/Q_{\left\{1,s\right\},r}^0=\left\{2.9,
  2.2\right\}$.
As expected, eigenmode analysis reveals the system to be totally
asymmetric, with the cavity coupling exclusively to the lower
waveguide (downwards propagation). The coupler is also observed to be
both binary and smooth, having no feature smaller than
$120$~nm. Nearly identical results are seen for the triply resonant
system (non-degenerate SFG) illustrated in Fig.~2 (b).  Moving to a
three-track cavity designed to support modes at
$\lambda_{\{1,2,s\}}=\{1500,\;907,\;565\}$~nm, with
$Q_{\{1,2,s\},r}^0=\left\{640,\;5.3\times 10^4,\; 3.2\times
  10^4\right\}$,
the algorithm is again able to realize critical coupling at all three
wavelengths, resulting in transmission $\lesssim 1\%$. Cavity
radiative lifetimes are also similarly enhanced, with
$Q_{\left\{1,2,s\right\},r}/Q_{\left\{1,2,s\right\},r}^0=\left\{2.2,\;1.7,\;3.1\right\}$.
In either inverse design, the coupling mechanism is found to be more
intricate than just the overlap of evanescent fields used for single
wavelengths. This is most pointedly seen in (2) and (3) of Fig.2 (b),
where over $99.5$\% of the energy density is in the cavity, yet
critical coupling occurs at over two wavelength of separation due to
the fields in the coupler. Moreover, for some cavity modes, even at a
single wavelength, it would not be possible to achieve critical
coupling using the evanescent tails of a waveguide
mode. For example, due to its tight confinement to the inner ring, even if the waveguide is made to touch the cavity, it is not possible to couple to the mode displayed in Fig. 2
(a.1) with better than $70\%$ efficiency. (Reducing the waveguide cross-section offers no improvement due to the creation of phase mismatch.)  Note that the radiative
quality factors of the cavities we have designed are smaller than
those typically considered for nonlinear processes. This choice was
made primarily to test the algorithm in cases involving dissimilar
waveguide and cavity mode profiles~\cite{xu2008silicon}. Nevertheless,
we note that for equivalent nonlinear performance, larger overlaps
$\beta$ and smaller radiative lifetimes are often preferable to higher
quality factors~\cite{lin2016cavity}.


As a final benchmark, Fig. 3 demonstrates a system attaining near
critical coupling at 6 frequencies (over an octave), a frequency comb
with large tooth spacing. (A more practical frequency comb coupler,
e.g. exhibiting critical coupling at over $100$ frequencies, will be
considered in future work.)  In this case, we begin with a
wavelength-scale ring resonator having unevenly distributed modes
$m=\{5\rightarrow 10\}$ at intervals
$\Delta f=\{0.101, 0.099, 0.096, 0.098, 0.096\}\mathrm{c}/1.5\mu$m,
with
$f_{m}=0.667c/\left(1.5\mu\text{m}\right)+\sum^{m-5}_{i=0} \Delta
f_{i}$. To simplify future comparisons, material dispersion is ignored
and the cavity, waveguide and coupler are all assumed to have a
constant permittivity of $\varepsilon=9.3514$.  As opposed to our
first two examples, where the modal resonance frequencies are fixed
constraints, adding a degree of modularity to the total system design,
here, we assume that the resonance frequencies must be tuned by the
coupler.  This conceptual shift requires some small modifications to
stabilize the optimization algorithm. Namely, we now include an
initial phase where the usual energy density objectives are replaced
by field overlap integrals with the eigenmodes of the cavity
$\int_{V}\text{d}V\;\text{Re}\left[\varepsilon(\omega_i,\mathbf{r})\mathbf{E}^{*}_{m}(\omega_i,\mathbf{r})\cdot\mathbf{E}(\omega_i,\mathbf{r})\right]$,
where the $m$ subscript denotes the $m^{th}$ mode of the bare cavity,
and, as before, the integration is restricted to lie within the
outermost material boundary of the cavity. The frequencies
$\left\{\omega_{i}\right\}$ where these computations are carried out
are initialized to match those of the bare cavity, and then slowly
transitioned to the desired resonances, i.e. an evenly distributed
set.  The converged output of this procedure is effectively a new
cavity having characteristics well-matched to the original coupling
optimization algorithm.  The coupler displayed in Fig.~3 achieves the
desired wavelength tuning and critical-coupling functionality.  The
resonance frequency intervals are equally distributed as
$\Delta f=0.098 \mathrm{c}/1.5 \mu$m, and good coupling (transmission
and reflection below $15\%$) is visibly present in each of the field
profile plots. Explicitly, the summed transmitted and reflected powers
of are found to be $\left\{2\%,~5\%,~13\%,~2\%,~1\%,~4\%\right\}$.
However, smaller minimal feature sizes, $\approx 15$~nm, and a larger
total footprint $4.5~\mu\text{m} \times 4.5~\mu\text{m}$ were required
to achieve these effects. The number of iterations needed for this
optimization was roughly the same as those of the previous examples,
leading to approximately linearly scaling of the total computation
time with the number of frequencies. 

In summary, we have shown that, in two dimensions, inverse design
provides a practical means of efficiently coupling light at multiple
widely separated wavelengths from a single channel (a waveguide) into
a compact, multimode cavity. Drawing from our recent work on the
design of compact microcavities for high-efficiency nonlinear
frequency conversion, we have successfully treated suggestive examples
for second-harmonic (SHG), sum-frequency (SFG), and frequency comb
generation (albeit for large tooth spacing). Critical coupling was
achieved, or nearly achieved, at all relevant wavelengths without
incorporating sharp components in the first two cases. In particular,
all features of the SHG system are larger than $120$~nm. Our results
continue the promising trend seen in application of inverse design to
free-space and on-chip couplers, rectifiers, and multiplexers,
indicating the potential of these techniques to enable significant
improvements in integrated nonlinear photonics.

\emph{Acknowledgments.---} This work was supported by the National
Science Foundation under Grant No. DMR-1454836, Grant No. DMR 1420541,
and Award EFMA-1640986; and the National Science and Research Council
of Canada under PDF-502958-2017.

%
\end{document}